\documentclass[twocolumn,apj]{emulateapj}

\shorttitle{Detection of supersonic downflows and associated heating events in the TR above sunspots}
\shortauthors{Kleint et al.}

\begin{document}

\title{Detection of supersonic downflows and associated heating events in the transition region above sunspots}

\author{L. Kleint\altaffilmark{1,2,3}, P. Antolin\altaffilmark{4}, H. Tian\altaffilmark{5},  P. Judge\altaffilmark{6}, P. Testa\altaffilmark{5}, B. De Pontieu\altaffilmark{2}, J. Mart\'{\i}nez-Sykora\altaffilmark{1,2},  K. K. Reeves\altaffilmark{5}, J. P. Wuelser\altaffilmark{2}, S. McKillop\altaffilmark{5}, S. Saar\altaffilmark{5},  M. Carlsson\altaffilmark{7}, P. Boerner\altaffilmark{2}, N. Hurlburt\altaffilmark{2}, J. Lemen\altaffilmark{2},   T. D. Tarbell\altaffilmark{2},  A. Title\altaffilmark{2}, L. Golub\altaffilmark{5}, V. Hansteen\altaffilmark{7},  S. Jaeggli\altaffilmark{8}, C. Kankelborg\altaffilmark{8} }
\altaffiltext{1}{Bay Area Environmental Research Institute, 625 2nd St, Ste. 209, Petaluma, CA 94952, USA; lucia.kleint@fhnw.ch}
\altaffiltext{2}{Lockheed Martin Solar and Astrophysics Laboratory, 3251 Hanover St., Org. ADBS, Bldg. 252, Palo Alto, CA  94304, USA}
\altaffiltext{3}{University of Applied Sciences and Arts Northwestern Switzerland, Bahnhofstrasse 6, 5210 Windisch, Switzerland}
\altaffiltext{4}{National Astronomical Observatory of Japan, 2-21-1 Osawa, Mitaka, Tokyo 181-8588, Japan}
\altaffiltext{5}{Harvard-Smithsonian Center for Astrophysics, 60 Garden Street, Cambridge, MA 02138, USA}
\altaffiltext{6}{High Altitude Observatory/ NCAR, P.O. Box 3000, Boulder CO 80307, USA}
\altaffiltext{7}{Institute of Theoretical Astrophysics, University of Oslo, P.O. Box 1029, Blindern, NO-0315 Oslo, Norway}
\altaffiltext{8}{Department of Physics, Montana State University, Bozeman, P.O. Box 173840, Bozeman, MT 59717, USA}


\begin{abstract}
IRIS data allow us to study the solar transition region (TR) with an unprecedented spatial resolution of 0.33\arcsec. On 2013 August 30, we observed bursts of high Doppler shifts suggesting strong supersonic downflows of up to 200~km/s and weaker, slightly slower upflows in the spectral lines \ion{Mg}{2} {\em h} and {\em k}, \ion{C}{2} 1336, \ion{Si}{4} 1394~\AA, and 1403~\AA, that are correlated with brightenings in the slitjaw images (SJIs). The bursty behavior lasts throughout the 2 hr observation, with average burst durations of about 20 s.
The locations of these short-lived events appear to be the umbral and penumbral footpoints of EUV loops. Fast apparent downflows are observed along these loops in the SJIs and in AIA, suggesting that the loops are thermally unstable. We interpret the observations as cool material falling from coronal heights, and especially coronal rain produced along the thermally unstable loops, which leads to an increase of intensity at the loop footpoints, probably indicating an increase of density and temperature in the TR.
The rain speeds are on the higher end of previously reported speeds for this phenomenon, and possibly higher than the free-fall velocity along the loops. On other observing days, similar bright dots are sometimes aligned into ribbons, resembling small flare ribbons. These observations provide a first insight into small-scale heating events in sunspots in the TR.
 \end{abstract}

\section{Introduction}

Energy is transported along coronal loops, which can be observed as siphon flows, coronal rain, or plasma accelerated by flares and explosive events. Early observations of the transition region (TR) reported bright fan-shaped features above sunspot umbrae, called plumes \citep{foukaletal1974}. Plumes are the bright lower parts of coronal loops and show downflows of 20--40 km/s at TR temperatures, with rare second components up to 150 km/s \citep{dere1982,nicolasetal1982,gurman1993}. Their main characteristic is enhanced emission at upper TR temperatures (10$^5$--10$^6$ K), supported by measurements of the differential emission measure \citep{brosius2005,tianetal2009}. Plumes can be stable over consecutive observing days \citep{brosiuswhite2004}, and fade when the downflows disappear \citep{brosius2005}. 

The velocities measured at the footpoints of coronal loops give us insight into the physical processes of energy transport and may help to answer the open question of the coronal energy budget. In particular, driving mechanisms, such as siphon flows, free-fall, or particle acceleration will result in different velocity distributions. The statistics of downflow velocities in the TR in a large number of sunspots are limited so far \citep{brekkeetal1990, gurman1993, brynildsenetal2004, tianetal2014} with some reports of no supersonic velocities above several umbrae, and others finding supersonic downflows in multiple sunspots, mostly above the penumbra.

\begin{figure*} 
   \centering 
   \includegraphics[width=.75\textwidth]{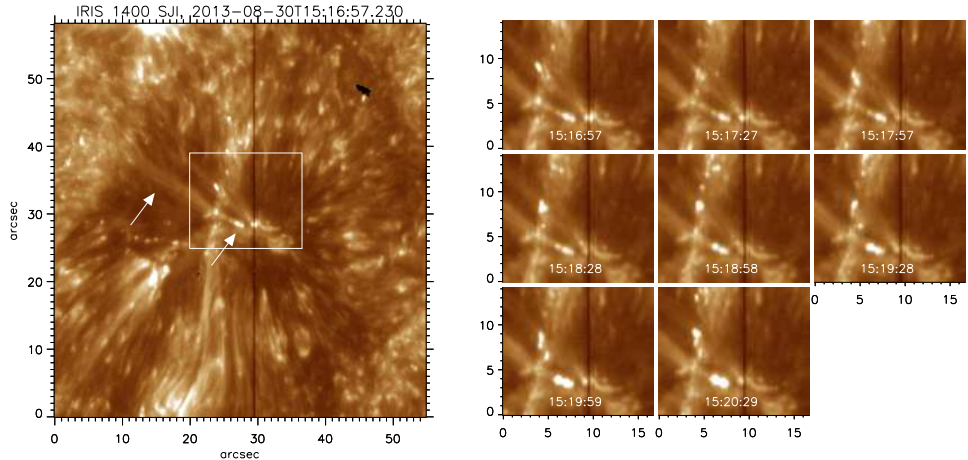}
   \caption{Sunspot with emission dots above its umbra/penumbra: AR 11836 with a lightbridge splitting the sunspot on 2013 August 30. A SJI 1400 overview is on the left, temporal evolutions of the region marked by a white box on the right. The vertical black line is the spectrograph slit. The left white arrow points to a faint loop that ends in a bright dot (right arrow). The images show that the emission evolves on timescales of minutes (see online movie).}
         \label{overview}
  \end{figure*}

\begin{figure*} 
   \centering 
    \includegraphics[width=.75\textwidth]{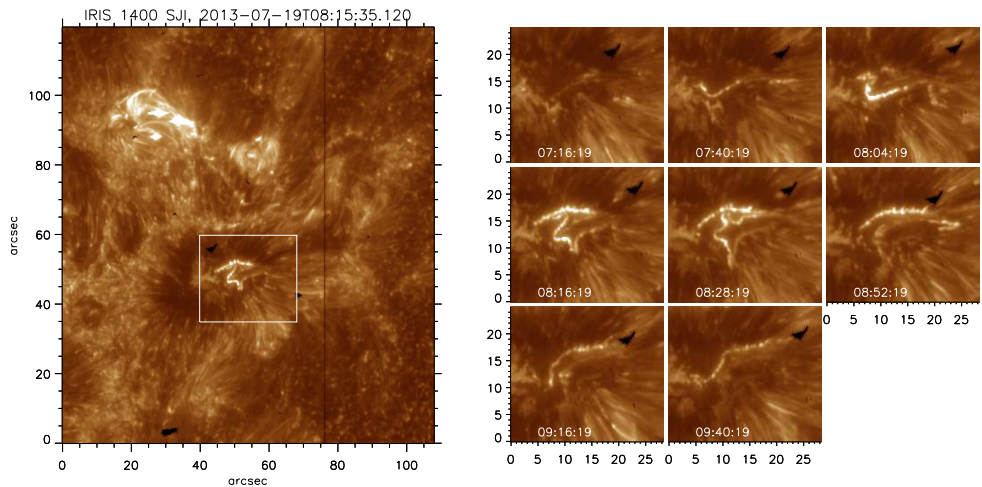}
   \caption{Similar to Figure~\ref{overview}. AR 11793 on 2013 July 19, which shows emission dots in the shape of ribbons that resemble strings of pearls (see movie).
   }
         \label{overview2}
  \end{figure*}

Coronal rain corresponds to emission in chromospheric and TR lines from neutral or partially ionized material occurring at time scales of minutes along coronal loops \citep{kawaguchi1970, leroy1972,levinewithbroe1977}. It is thought to be produced by the thermal instability mechanism \citep{parker1953,field1965} when radiative losses exceed heating. Strong heating at loop footpoints may lead to a high plasma density in loops and therefore strong radiative losses. Numerical simulations indicate that in specific physical conditions (depending on the loop length and the heating scale length, among others), the conductive heating flux cannot compensate the radiative losses in the corona \citep{antiochosetal1999,karpenetal2001}. The radiative losses increase as the plasma cools, which may result in catastrophic (self-amplifying) cooling. The loss of temperature is accompanied by a loss of pressure, which accretes plasma from the surroundings, leading to the localized formation of condensations \citep{goldsmith1971,hildner1974,antiochosklimchuk1991,mulleretal2004}. Such structures appear clumpy in chromospheric lines and are usually observed to fall much slower than free fall \citep{mackaygalsgaard2001, antolinverwichte2011}.

In the photosphere and chromosphere, supersonic downflows are either relatively rare, or hard to observe. They were reported in the \ion{He}{1} 10830 \AA\ line  at 40 km/s in plage and at loop footpoints near a pore \citep{schmidtetal2000,laggetal2007}. In the latter case, the photospheric velocity at that location was only 1 km/s, indicating a transition from supersonic to subsonic velocities between those layers. Such a transition leads to a shock front in the atmosphere and is visible as emission \citep{cargillpriest1980}.

In this Letter, we will focus on first results from the recently launched Interface Region Imaging Spectrograph \citep[IRIS][]{iris2014} showing highly dynamic bright footpoints of coronal loops/plumes with supersonic velocity components.


\begin{figure*} 
   \centering
       \includegraphics[width=.9\textwidth]{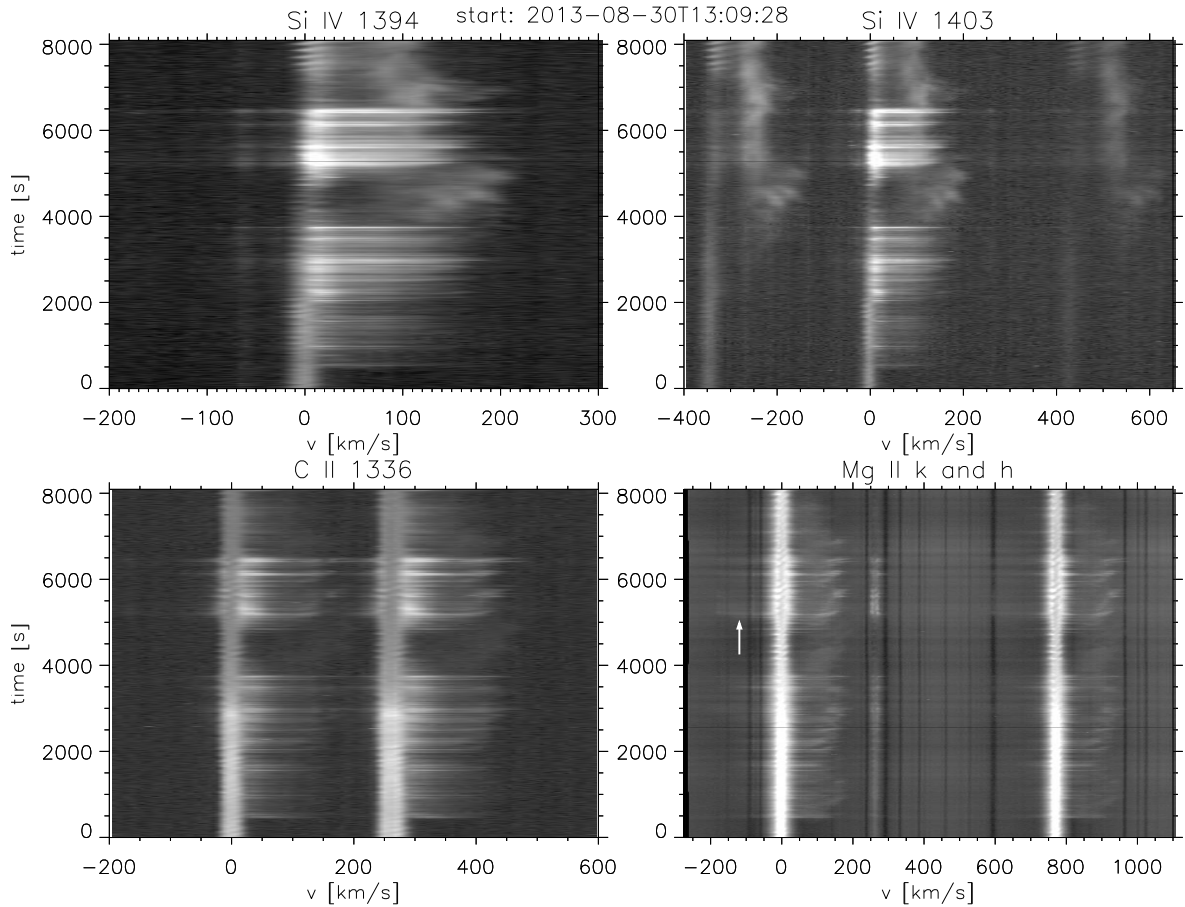}
  \caption{Temporal evolution (vertical axis) of a dot that repeatedly shows emission in four different spectral windows from 13:09 to 15:24 UT on 2013 August 30. Horizontal axes are labeled in Doppler units with zero chosen to coincide with one spectral line. Strong downflows (up to 200 km/s) and weaker upflows (example location denoted by the arrow) are visible with burst lasting about 20 s. Also visible are sunspot oscillations with their characteristic sawtooth pattern. The images were scaled with the 10th root to strongly increase their contrast.}         \label{lambdat}
  \end{figure*}

\section{Observations and data reduction}
\label{obs}

IRIS records spectra in three passbands: the near ultraviolet (NUV, 2782.7--2835.1 \AA), and the far ultraviolet (FUV, 1331.7--1358.4 \AA\ and 1389.0--1407.0 \AA). Additionally, a slitjaw image (SJI) can be obtained simultaneously with the spectra, in one of four different filters.

For the velocity analysis, we focus on an observation taken on 2013 August 30, 2013 from 13:09 -- 15:24 UT. This observation was a medium-sized (60\arcsec\ along the slit) sit-and-stare with an exposure time of 4 s, targeting AR~11836. SJI were obtained in the 1400 (\ion{Si}{4}) and 2796 (\ion{Mg}{2} k) filters with a cadence of 10 s. Their spatial resolution is 0.33\arcsec\ and 0.4\arcsec, respectively. Additionally, we show SJI 1400 images of AR 11793 that were obtained on 2013 July 19 during various observing sequences from 6 to 10 UT to illustrate the fast changes of a ribbon of bright emission dots. Calibrated Level 2 data were used, which include a dark correction, flatfielding, geometric and wavelength corrections. No detectable flares occurred during our observations.

We observed highly dynamic bright points in sunspots, which can change their appearance within seconds to minutes. They appear above locations in the umbra and penumbra, as determined from photospheric images. Figures~\ref{overview} and \ref{overview2} show examples from two observing days. For the global context, we aligned the 1400 IRIS SJI with the 1600 \AA\ passband of SDO/AIA \citep{lemenetal2012}. The co-alignment shows that the bright dots coincide with footpoints of coronal loops and is discussed in Section~\ref{analysis}.

We found these bright dots to be relatively common in active regions, but they do not appear in every sunspot. However, due to their fast changing nature and their small size, it has been rare so far to have the IRIS spectrograph slit point exactly through one of them. The observation from 2013 August 30 lasted more than 2 hr and captured the spectra of several dots. Unfortunately, the slit did not cross any dots on 2013 July 19 and we thus have no information on the Doppler shifts, or confirmation that the dots in fact are the same phenomenon, even though their sizes and morphologies are similar. The temporal evolution on 2013 August 30 clearly shows apparent plasma motions into the sunspot along loop trajectories with the footpoints repeatedly increasing their intensities. While the events in the umbra are easiest to observe due to their contrast, similar events seem to occur in the lightbridge and the penumbra, but we do not exclude other mechanisms for other bright dots, especially the fainter dots in the penumbra.

\begin{figure*} 
   \centering
       \includegraphics[width=.4\textwidth]{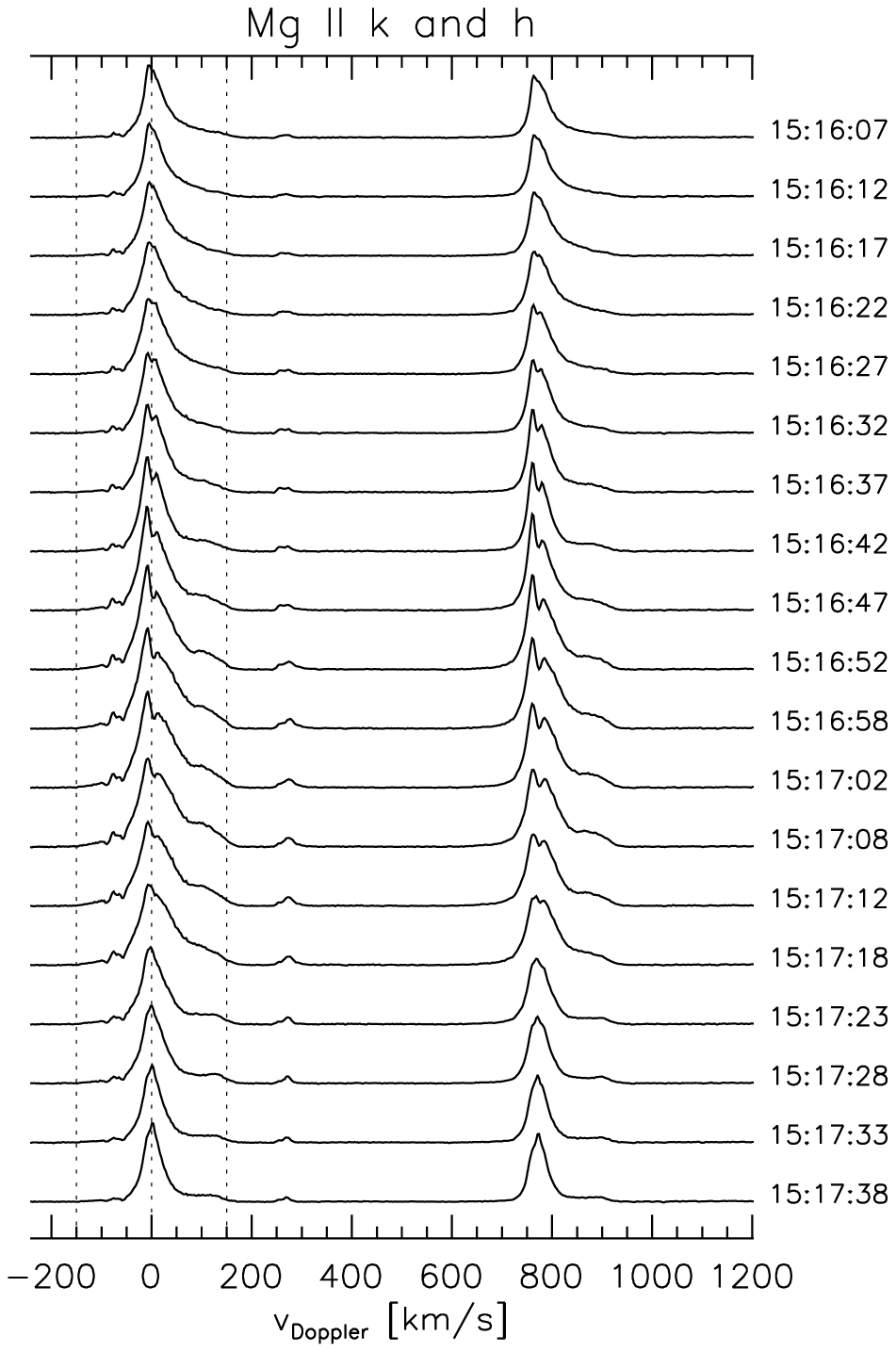}
     \includegraphics[width=.4\textwidth]{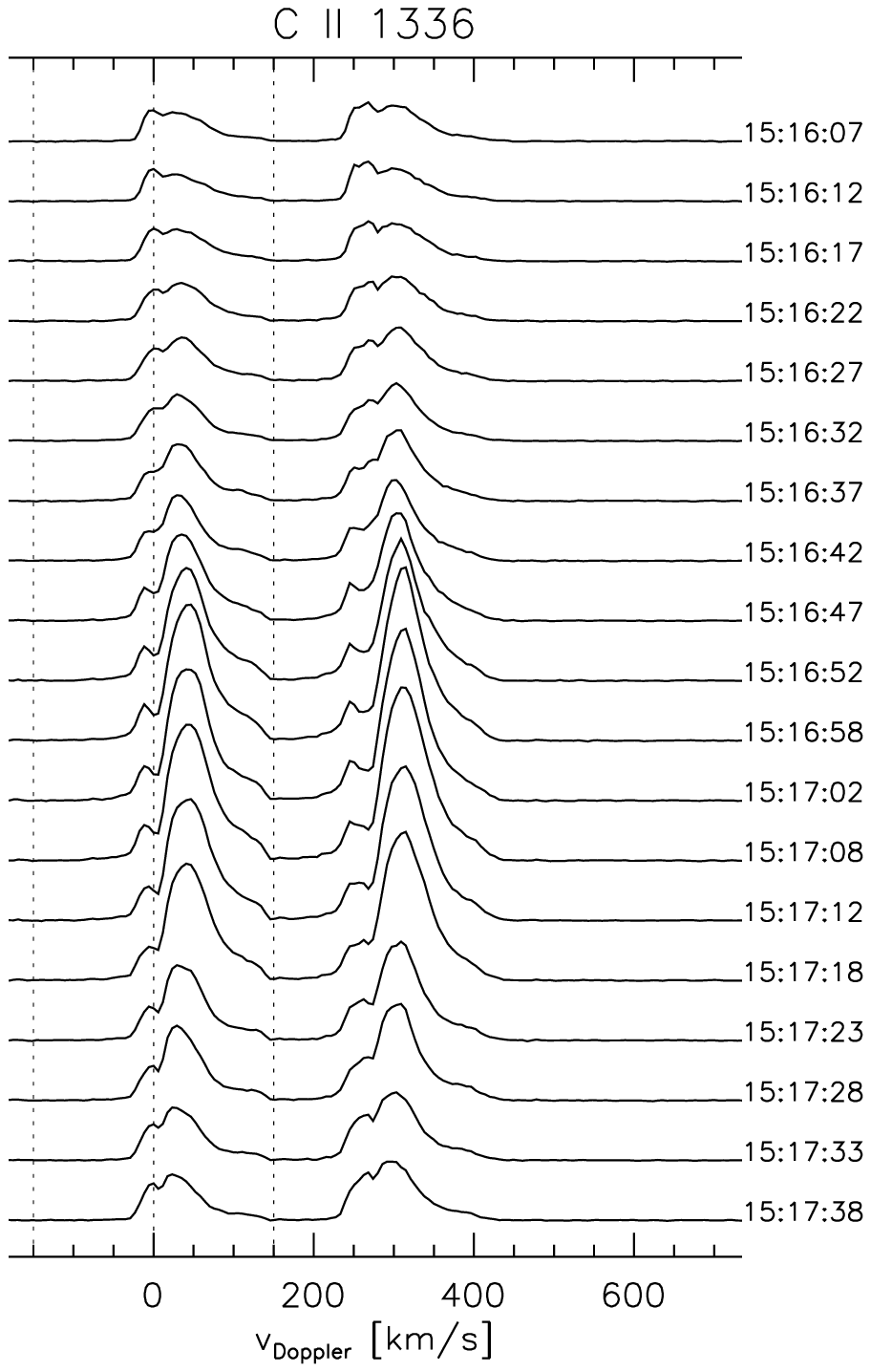}
        \caption{Evolution of \ion{Mg}{2} and \ion{C}{2} 1336 spectra during the appearance of a bright dot in the umbra on 2013 August 30. Their scaling is arbitrary and different for \ion{Mg}{2} and \ion{C}{2}. Especially the FUV spectra, such as \ion{C}{2}, show significant changes in intensity and line shapes.}
         \label{evolnuv}
  \end{figure*}

\section{Properties of the bright footpoints}\label{analysis}

The  drastic intensity changes evolve within seconds to minutes, and may appear anywhere in sunspots. The shape of the ribbons may also change rapidly. Some observations only show single dots, which seem to be common in light bridges and the penumbra. With a full-width at half maximum of 0.35\arcsec - 0.7\arcsec\ (2 -- 4 IRIS pixels), they are too small to be resolved in AIA images, which is why these events have not been reported before.

Figure~\ref{lambdat} shows the temporal evolution of the Doppler shifts of several spectral lines in a pixel inside the sunspot's umbra where a bright dot appears at times. Bursts of strong downflows (redshifts) up to 200 km/s are visible in all spectral lines. Significantly weaker intensity enhancements (``blueshifts''), generally below --100 km/s, are also visible in all lines, but best in \ion{Mg}{2}, probably because its highest signal-to-noise ratio.  They could be either true upflows, changes of the line profile shapes (line broadening or raised line cores), or downflows from other spectral lines. 
The observed ``blueshifts'' are not correlated to the strongest line intensity increases, which would favor a thermal broadening interpretation. Because they are visible in optically thin and optically thick lines, line shape changes are also unlikely. Redshifts from other spectral lines are unlikely because there is no line visible to which the shifts could be attributed. The most likely explanation is therefore that they are of nonthermal origin, for example, upflows or turbulence, which coincide temporally with strong downflows. However, because of their low signal-to-noise ratios, it cannot be determined which spectral line shows the largest blueshifts. In the bottom right panel, there is a line visible at x=260 km/s, which belongs to the 3p--3d triplet of \ion{Mg}{2} and also sometimes shows emission, generally during the downflow events. The observed downflows are strongly supersonic, as the sound speed ($v_s$) lies below 50 km/s, assuming the adiabatic approximation $v_s = \sqrt{\gamma R T}$ and a temperature around 10$^4$ -- 10$^5$ K. There is no time lag for these events between the different spectral lines, at least to the accuracy of the interval of 5 s between consecutive spectra.

The line profiles change their shapes significantly and on short timescales. Figure~\ref{evolnuv} shows the evolution of the \ion{Mg}{2} and the \ion{C}{2} 1336 spectra during 1.5 minutes (top to bottom), which coincided with a bright dot appearing in the SJI. The vertical dotted lines denote $\pm$150 km/s for reference. A sudden increase of intensity with significant downflows (extra peak in all line profiles around x=100 km/s) is visible. The FUV lines generally exhibit more drastic changes in intensity than \ion{Mg}{2}. All these high-velocity events appear as additional, shifted line components, meaning that some plasma in the field of view is at rest. Because of the multiple components and strongly variable line profiles, Gaussian fitting has proven too unreliable for these observations.

\begin{figure*} 
   \centering 
    \includegraphics[width=\textwidth]{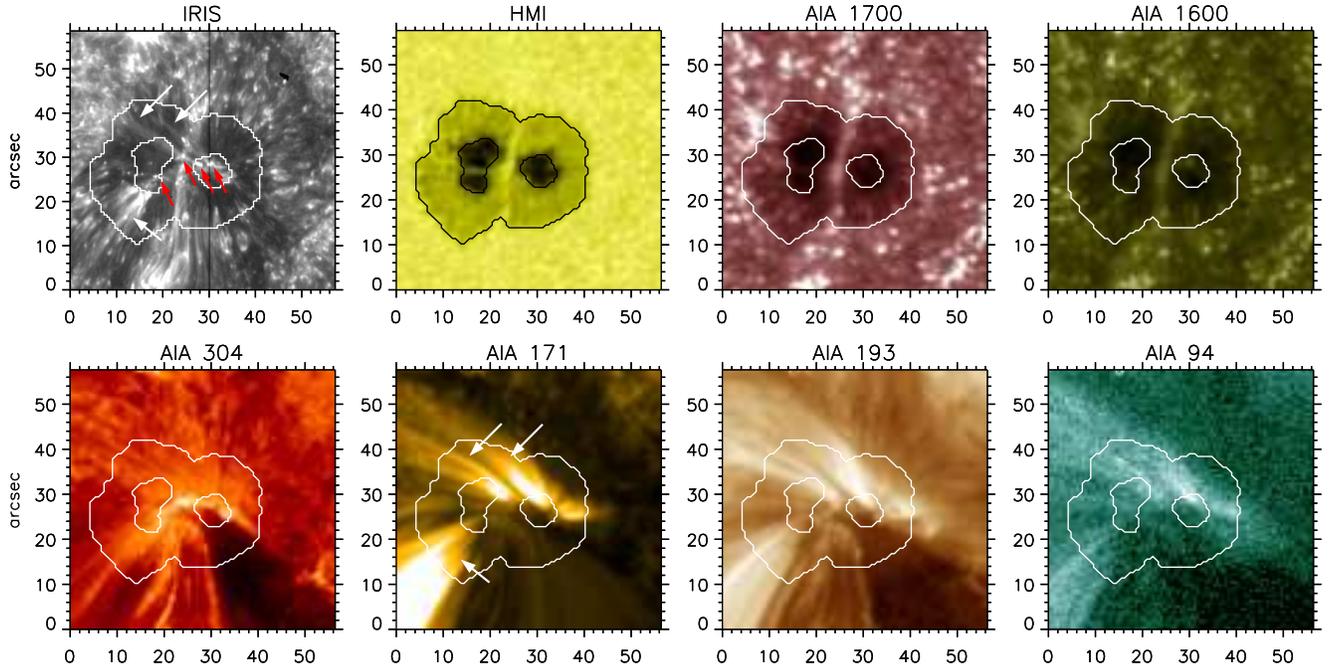}
   \caption{IRIS SJI 1400 (top left) and SDO images at the same time (15:16 UT on 2013 August 30). Faint loop parts can be seen in the IRIS image,  which correspond to bright loops in AIA 171 and 193 (white arrows). The bright dots visible in IRIS images are located at the ends of these loops (small red arrows). See movie.}
         \label{aia}
  \end{figure*}

Figure~\ref{aia} shows AR 11836 in several SDO passbands, ranging from photospheric to coronal temperatures with the IRIS reference image at the top left. Traces of coronal loops can be seen in the IRIS image ending in bright dots in the umbra. The faint IRIS loops correspond to bright loops in AIA 171, 193 and 94, whose intensities vary (see online movie). Apparent flows along these coronal loops into the sunspot can also be observed in these AIA passbands, and especially in the 304 passband, where the loop system appears darker. A decrease in EUV intensity can be caused by absorption of cool material \citep{anzerheinzel2005,landireale2013}.

\section{Discussion}
Visually, the plasma falling along the loops observed with IRIS closely resembles coronal rain. Ground-based observations have found that coronal rain has an average speed of 60--70 km/s, with peaks up to 140 km/s, and varies on timescales of minutes \citep{antolinluc2012, antolinetal2012}. It is estimated that 7\%--30\% of the coronal volume exhibits rain, although estimates are difficult mainly because of the low contrast of coronal rain on disk. To our knowledge, this may be the first observation tracing coronal rain into a sunspot's umbra. One question is why our observed velocities are slightly higher than those reported previously. It could simply be an effect of different spatial resolutions, or it could be related to the fact that we observe events inside the umbra, where the gas pressure is lower, which may lead to less deceleration. Siphon flows may play a role in the observations, but it is unclear to what extent, especially because the other loop footpoint is not observed. However, the process of coronal rain formation involves flows, which does not exclude siphon flows. Because falling dark blobs can be observed in IRIS and AIA images, coronal rain is very likely the dominating process. Visual tracing of blobs in IRIS data, by creating intensity evolutions along the blob trajectories, reveals much slower speeds (below 50 km/s) than the IRIS spectra. The ratio of projected to line-of-sight (LOS) velocity components is however similar to that observed for off-limb observations of rain falling almost perpendicularly to the LOS. In that case, the projected component is the faster component. Using Equations (2)-(4) from \citet{antolinetal2012} adapted to an on-disk scenario, we can determine that the observed projected velocity of the blobs corresponds to an angle of $\sim$5--20 deg above the footpoint with respect to the loop baseline (angle $\varphi$ in their Figure~12). Our observations also resemble those of \citet{realeetal2013}, although on a much smaller spatial scale.

\subsection{Velocities - Faster than Free Fall?}
To investigate the process of energy transport, we calculated the free fall velocity along elliptical loops according to \citet{antolinverwichte2011}. For an assumption of zero initial velocity at the loop top ($s=0$), the velocity $v_s$ along the loop  can be written as $v_s(s) = \sqrt{2 \int_0^s g_{sun} \cos{\theta}(s') ds'}$. From AIA images, we determined the loop semilength (half the linear distance between its footpoints, $a$) to be about 50 Mm, while its height ($h$) is a free parameter and cannot be determined from the images. The maximum velocities for a blob in such an elliptical loop vary between 145 km/s and 180 km/s, for $h/a$ ratios of 0.5 -- 2, respectively. Under pure free-fall conditions, we would therefore need initial velocities of 20 -- 50 km/s in order to match the observed values. The required initial velocity may even be higher if some expected deceleration due to collisions with plasma at rest occurs. Flows with such values have been previously observed near the apex of the loops with Hinode/SOT \citep{ofmanwang2008,antolinetal2010}.

\subsection{Similarity to Flare Ribbons}
Thermal instability appears to be the same for post-flare loops and non-flaring active regions \citep{foukal1978,schmiederetal1995}. The observation on 2013 July 19 showed slightly more activity than 2013 August 30, but there were only microflares, the strongest of which occurred toward the end of the observations while the emission dots were fading (see online movie). Coronal rain was visible in AIA images during the phase of the simple ribbon (before 7.40 UT). Then, a jet-like activity ejected plasma, which can be seen falling back down in AIA 304, additionally to the coronal rain, and which coincides with the time when the ribbon changes shape. A visual resemblance to flare ribbons is striking. They show similar small-scale brightenings \citep[e.g.,][]{kleint2012} when observed with sub-arcsecond resolution. This is in agreement with the standard flare model, where particles are accelerated along magnetic field lines into the TR and chromosphere, probably leading to isolated brightenings when the energy between neighboring loops is not equal. The precipitating particles heat the footpoints, leading to evaporation of dense and hot material into coronal loops. As the radiative losses outweigh the heating, a thermal instability may form in post-flare loops, which then may lead to coronal rain falling into the ribbons and possibly contributes to their brightness. Ribbon-like structures suggest a heating correlation between neighboring loops. Such a correlation has also been observed by \citet{antolinluc2012} for coronal rain falling into pores, but with smaller correlation lengths of about 2 Mm. Similar correlations are obtained in 2.5D simulations \citep{fangetal2013}, and can be explained by a similar thermodynamic evolution of neighboring loops due to a similar geometry and heating properties. On 2013 July 19, the coronal rain and the falling plasma from the jet may both have contributed to the dynamics of the ribbon, whose emission dots closely resemble those from 2013 August 30.

We verified that no bright dots or unusually high velocities were visible in the photospheric Hinode/SP 6301/6302 raster, taken within an hour of our observations on 2013 August 30. The fact that there were no flares during the IRIS observations may explain why the bright dots were constrained to higher atmospheric layers as the energy transported along the loop was dissipated before reaching the photosphere and it would be interesting to capture and compare an event connected to a flare. 

\subsection{Local Heating in the TR}
 Emission dots may either be observed when the supersonic falling plasma undergoes a shock to subsonic speeds  \citep{cargillpriest1980} or possibly by collisions of the plasma with the more dense atmosphere \citep{realeetal2013}. While the location of the shock front cannot be determined here, it must be lower than the formation height of \ion{Mg}{2}, otherwise there would be no supersonic velocities, and higher than that of \ion{Fe}{1} 6301/6302, because no unusually high velocities were seen there.
 
 Probably because of the high downflow velocities, there is no observable time lag between the different observed spectral lines that form between 10$^4$ and 10$^5$ K. For a rapidly decreasing temperature due to non-equilibrium, and with the possibility that the loops are composed of different strands responding slightly different to the heating and cooling, it is very likely to obtain multi-temperature structures along the LOS \citep{realeetal2012apj}. But since they investigated cooling on timescales of minutes, and not seconds as in our case, future modeling efforts should verify that this process may occur on these shorter timescales.
 
 An increase of the local density and temperature is very likely, both in the shock, and the collision scenario. The observation of coronal rain also indicates heating at the loop footpoints, as the current models for coronal rain require a thermal instability. The observed upflow may be explained similarly to ``chromospheric'' (in this case TR) evaporation: the TR heats faster than radiation can carry the energy away, the temperature increase leads to an increase of pressure and to an upward force, sending plasma back into the corona. It is unclear if these upflows can flow along the same field line as the downflows and one may need to consider the role of partial ionization effects. Small scale brightenings with similar timescales have been observed in moss \citep{testaetal2013} and near active regions \citep{regnieretal2014}. However, the mechanisms seem to be different in those cases because no coronal rain was observed among other differences.

\section{Conclusions}\label{results}

We reported small scale brightenings that appear as bright dots or ribbons in umbrae and penumbrae of sunspots. Spectra of several lines that form at temperatures of 10$^4$--10$^5$ K show bursts of supersonic Dopplershifts, mostly downflows up to 200 km/s, but also weaker upflows on 2013 August 30. Simultaneous AIA images of coronal passbands reveal these dots to be located at the ends of coronal loops.

The maximum observed velocities probably require an initial velocity at the loop apex, e.g.\,are faster than free fall. Our comparably high final velocities at the footpoints could be related to the lower gas pressure in sunspots or an unknown acceleration mechanism. Once the downflowing plasma hits more dense atmospheric layers, heating will be caused through collisions or shocks when the speeds go from supersonic to subsonic. 

The kinetic energy from the downflowing plasma is probably dissipated partly as local heating in the TR, which may lead to the observed bright dots. The duration of a burst (intensity increase and large Doppler shifts) is about 20 s. The IRIS SJI show blobs of material falling into the sunspot, which may be the first observation tracing coronal rain into an umbra. These observations may contribute to the solution of the coronal energy budget and in the future, it should be investigated how frequent these events are, how much energy is deposited, and through simultaneous observations in photospheric and chromospheric lines, at which layer the energy is deposited. Our observations suggest the presence of heating events in the TR above sunspots, probably leading to catastrophic cooling of loops. The successive brightenings resembling a string of pearls suggest the existence of a strong and rather peculiar correlation in heating for neighboring loops rooted in sunspots. These observations demonstrate the power of IRIS to possibly resolve the fine structure of downflows in the TR and they will help to constrain related modeling efforts in terms of up- and downflow velocities, heating leading to catastrophic cooling, and heating in the low atmosphere due to collisions.

\acknowledgments
IRIS is a NASA small explorer mission developed and operated by LMSAL with mission operations executed at NASA Ames Research center and major contributions to downlink communications funded by the Norwegian Space Center (NSC, Norway) through an ESA PRODEX contract.

\bibliographystyle{apj}
\bibliography{journals,ibisflare}

\end{document}